\documentstyle[twoside,fleqn,espcrc2,epsf]{article}

\newcommand{\AmS}{{\protect\the\textfont2
  A\kern-.1667em\lower.5ex\hbox{M}\kern-.125emS}}

\title{ Numerical studies of confinement in the Landau gauge\\
by the larger lattice simulation}

\author{Hideo Nakajima\thanks{e-mail nakajima@is.utsunomiya-u.ac.jp}\\ 
Department of Information Science, Utsunomiya University, 321-8585 Japan \\
Sadataka Furui \thanks{e-mail furui@liberal.umb.teikyo-u.ac.jp}\\
School of Science and Engineering, Teikyo University, 320-8551 Japan}

\begin{document}

\begin{abstract}
The infrared behaviour of QCD in the Landau gauge is studied in $24^4$
and $32^4$ lattices by measuring the Kugo-Ojima parameter, gluon propagator,
ghost propagator and the QCD running coupling.
The Kugo-Ojima parameter c remains about 0.8 instead of the expectation value
1. We observe infrared suppression of the
singularity of the gluon propagator and enhancement of the ghost propagator but
the shifts are smaller than those of Dyson-Schwinger approach. The momentum dependence of the running coupling in the infrared region is
weak and $\alpha_s(0)\simeq 1$.   
\vspace{1pc}
\end{abstract}

\maketitle

\section{The infrared behaviour of QCD in the Landau gauge}

The colour confinement in QCD can be formulated as an absence of coloured asymptotic states in the BRST formalism, and the presence of long-range $1/p^4$ correlation between colour sources. In the Gribov-Zwanziger's theory of Landau gauge QCD\cite{Zw}, the long-range correlation is attributed to the ghost propagator, i.e. the inverse of the Faddeev-Popov(FP) matrix. The ghost propagator is more singular and the gluon propagator is less singular than the tree level in the infrared region. 

In Kugo-Ojima's theory\cite{Ku}, a sufficient condition for an absence of coloured asymptotic states is that the parameter $u^{ab}$ defined by the two-point
function of  the FP ghost fields, $c(x),\bar c(y)$,
and $A_\nu(y)$\cite{NFY}, satisfies $u^{ab}(0)=-\delta^{ab}$.

\section{The Kugo-Ojima confinement criterion and the Gribov-Zwanziger's theory}
\begin{table*}[htb]
\label{table:1}\caption{The Kugo-Ojima parameter $c$ ,trace $e/d$ and the horizon function deviation factor $h$ in  $U-$linear(suffix 1) and $\log U$(suffix 2)  version, $\beta=6.0$. The bottom line is of $\beta=6.4$.}
\newcommand{\m}{\hphantom{$-$}}
\newcommand{\cc}[1]{\multicolumn{1}{c}{#1}}
\renewcommand{\tabcolsep}{1.2pc} 
\renewcommand{\arraystretch}{1.1} 
\begin{tabular}{@{}ll|lll|lll}
\hline
& $L$ &$c_1$ & $e_1/d$ & $h_1$ & $c_2$ & $e_2/d$ & $h_2$ \\
\hline
&16&  0.576(79) &   0.860(1) & -0.28 & 0.628(94)& 0.943(1) & -0.32\\
&24&  0.695(63)  &  0.861(1) & -0.17 & 0.774(76)& 0.944(1) & -0.17\\
&32&  0.706(39)  &  0.862(1) & -0.15 & 0.777(46)& 0.944(1) & -0.16\\
\hline
&32& 0.650(39) & 0.883(1) & -0.23 & 0.700(42)& 0.953(1) & -0.25\\
\hline
\end{tabular}

\end{table*}

Let two-point tensor ${G_{\mu\nu xy}}^{ab}$ be
\begin{equation}
{G_{\mu\nu xy}}^{ab}=
 {\rm tr}\left({\lambda^a}^{\dag}
D_\mu \displaystyle{1\over -\partial D}(-D_\nu)\lambda^b\right)_{xy}.
\label{eqq3}
\end{equation}
and the horizon function $H(U)$ 
\begin{equation}
H(U)=\sum_{x,y,a}{G_{\mu\mu xy}}^{aa}-(N^2-1)E(U)
\end{equation}
where $N$ is the number of colours, now $N=3$.
 $E(U)$ depends on the options of the definition of gauge field, $U$-linear or 
$\log U$ type;

$U$-linear: $E(U)=\sum_l{1\over N}{\rm Re}\ {\rm tr} U_l,$ 

$\log U$: $E(U)={\sum_{l,a}}{\rm tr}\left ({\lambda^a}^\dag S({\cal A}_l)\lambda^a\right)/(N^2-1)$,
where ${\cal A}_l=adj_{A_l}$, and 
$S(x)={{x/2\over {\rm th}(x/2)}}$ .

One can show that
in the core region\cite{Zw}, the horizon function $H(U)$
is negative and in the infinite volume limit
$\lim_{V\to \infty}{\langle H(U)\rangle/ V}=0$
which is called  horizon condition.

Zwanziger showed  in his lattice formulation 
that the ghost self-energy cancels precisely the zero-th order $p^2$ 
term and the ghost propagator behaves as $1/p^4$ in the
infrared asymptotic region 
provided no additional non-perturbative effect is involved
\cite{Zw}. Recent results of Dyson-Schwinger approach suggest the nonperturbative aspects are in effect.

Kugo also showed in the continuum formulation that 
the ghost propagator $G(p^2)/p^2$ behaves in the infrared asymptotic region
as\cite{Ku}
\begin{equation}
G(p^2)/p^2\sim {1/[p^2(1+u(0)+O(p^2))]}
\end{equation}
where $u(p^2)^{ab}=\delta^{ab}u(p^2)$. Thus Kugo-Ojima condition leads
to the same singular behavior as Zwanziger's, and derives the universal
renormalizaiton factors as
$\displaystyle {Z_1\over Z_3}={1\over \tilde Z_3}=1+u(0)=0.$

Putting Kugo-Ojima parameter as 
$u(0)=-c,$
one finds that the horizon condition is
written as
$\displaystyle\left({e\over d}\right)+(d-1)c-e=(d-1)\left(c-{e\over d}\right)=(d-1)h=0. $
where $e=E/V$ and $d=4$.

Thus one sees that both the above arguments appear to be consistent
with each other provided the lattice covariant derivative meets with 
the continuum one in the naive sense, i.e., ${e/d}=1$.

Kugo discussed in his lecture\cite{Ku} that the colour confinement can be
realized via either
(1) $Z_1=0$ and $Z_3=$finite or (2) $Z_1/Z_3=0$ but/and $Z_3=0$

Zwanziger\cite{Zw,Zw1} argues that the second possibility is realized in the SU(2)
colour case.

Our numerical results are summarized in Table 1.
We obtained that $u^a_b(0)$ is consistent to $-c\delta^a_b,\ c=0.8$ in
$SU(3)$ quenched simulation, $\beta =6.0$, on $24^4$ and $32^4$.

In the case of $\beta =6.4$ on $32^4$, the value of $e/d$ becomes closer to 1, but $|h|$ becomes larger as compared to those of $\beta=6.0$, and thus $c$ becomes smaller. We need to perform larger lattice simulation to approach the continuum limit.

\section{The two-point functions (gluon propagator, ghost propagator)}

We measured gluon propagator of $\beta=6.0$ on lattice $16^4$, $24^4$ and $32^4$, and observed clear tendency of the infrared suppression in the case of $24^4$ and $32^4$. 
\begin{figure}[htb]
\renewcommand{\tabcolsep}{4pc} 
\begin{center}
\epsfysize=100pt\epsfbox{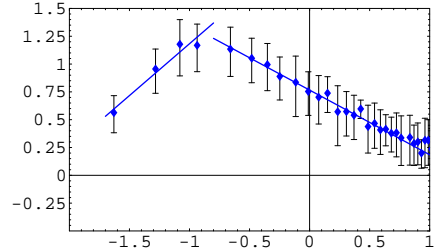}
\caption{The log of the gluon dressing function as a function of the log of the lattice momentum $pa$. $\beta=6.0$, $32^4$ in $U$-linear version. }
\label{gl32}
\end{center}
\end{figure}

\begin{figure*}[htb]
\begin{center}
\begin{minipage}[htb]{0.47\linewidth} 
\begin{center}
\epsfysize=100pt\epsfbox{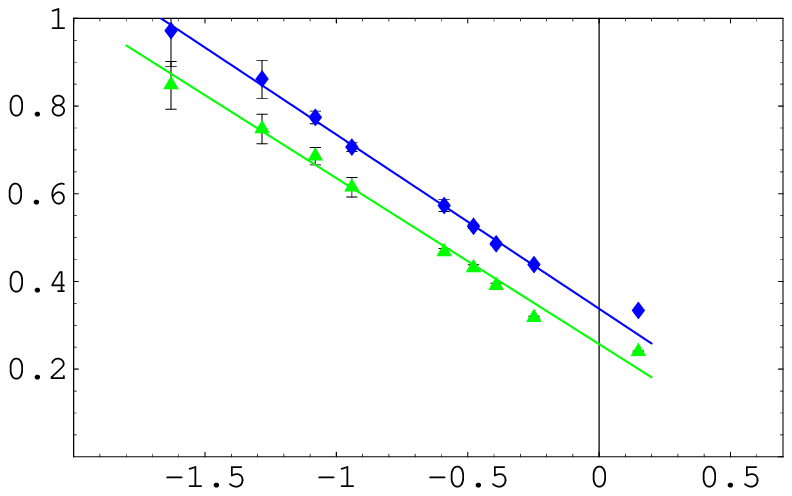}
\label{gh32l}
\end{center}
\end{minipage}
\begin{minipage}[htb]{0.47\linewidth} 
\begin{center}
\epsfysize=100pt\epsfbox{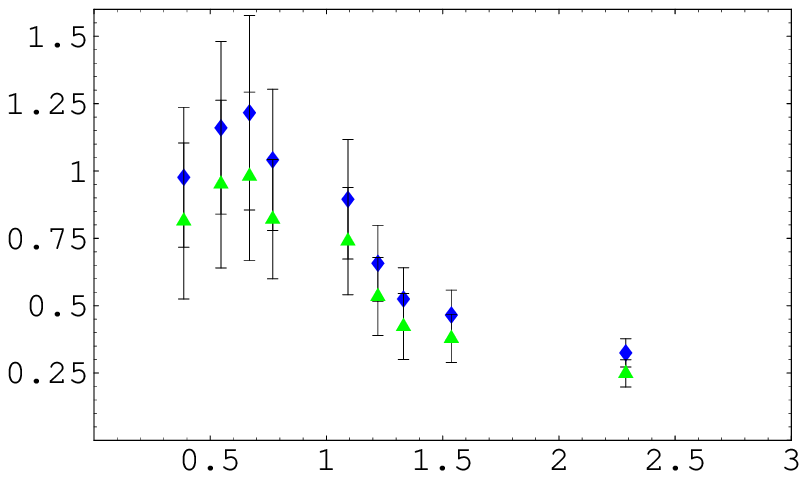}
\label{alp32}
\end{center}
\end{minipage}
\caption{The log of the ghost dressing function as a function of the log of the lattice momentum $pa$(left) and the running coupling $\alpha_s$ as a function of the lattice momentum $p$(GeV)(right). Triangles are in $\log U$ and diamonds are in $U$-linear version of $\beta=6.0$, $32^4$ lattice. }
\end{center}
\hfil
\end{figure*}
The ghost propagator is infrared divergent and its singularity can be
parametrised as $\displaystyle p^{-2(1+\alpha_G)}$, where $\displaystyle p^2=\sum_{\mu} (4 \sin^2{ \pi k_\mu\over L})$, $(-L/2 <k_\mu\le L/2)$. 
In the $32^4$ lattice, the overall scale difference in the $\log U$ and $U-$linear versions disappear in gluon propagator, but it still remains in ghost
propagator.

The gluon propagator and the ghost propagator are evaluated also from
Dyson-Schwinger approach\cite{FAR}. They found qualitatively
similar solutions for 
dressing functions of the gluon and the ghost,
$Z(p^2)\sim (p^2)^{-\alpha_D}$ and
$G(p^2)\sim (p^2)^{-\alpha_G}$
with $\alpha_D+2\alpha_G=0$, $\alpha_G\simeq 0.595$ and
$\alpha_D\simeq -1.19$\cite{FAR}.

The gluon propagator vanishes at $0$ momentum if $\alpha_G>0.5$. 
Our lattice simulation of the ghost dressing function
$G(p^2)$ shows that the value $\alpha_G\sim 0.18$ remains unchanged 
at $1/3$ of the Dyson-Schwinger one through $16^4$ 
to $32^4$ simulations.
The negative $\alpha_D$ appears in $24^4$ lattice and 
$\alpha_D\simeq -0.4$ in $32^4$ lattice. 
We observe the tendency that $|\alpha_D|$ increases as the lattice size 
increases, but our data still stays about $1/3$ of the prediction of the 
Dyson-Schwinger approach.

In contrast to the first data of the ghost propagator\cite{SS}, 
suppression at the lowest momentum is not seen in our data of the ghost 
propagator. For calculation of the ghost propagator,
we adopted the multigrid Poisson solver 
whose accuracy was kept within $10^{-5}$,
and we set 1\% as an ending condition of perturbative calculation of the
inverse FP operator. The Landau gauge accuracy of $div A=0$ is
$10^{-4}$ in maximum norm.

Using the ghost and gluon dressing function, the running coupling can be evaluated from $\alpha_s(pa)=(g^2(a)/4\pi) Z(pa) G(pa)^2\sim (pa)^{-2\alpha_D-4\alpha_G}$, where $a$ is the lattice spacing. In the case of $\beta=6, 32^4$, which has the smallest $|h|$, the suppression of the gluon dressing function due to the loss of correlation of the gluon field and the enhancement of the ghost dressing function cancel. We find  $\alpha_D+2\alpha_G\simeq 0$ near the lowest momentum point and $\alpha_s(0)\simeq 1$. The infrared suppression of the running coupling measured from the three point function of gluon\cite{Orsay} and from the ghost-antighost gluon\cite{FN} in $\overline{MOM}$ scheme would be attributed to the loss of correlation of gluon fields.  

\section{Discussion and outlook}
In the $\beta=6, 32^4$ lattice simulation $|h|$ is about 0.15 and we need extrapolation to $|h|=0$. There remain finite size effects in ghost propagator in $32^4$ lattice, and also in the Kugo-Ojima parameter which manifest themselves in their dependence
on the definition of the gauge field. The continuum limit should not depend on the definition of the gauge field and so a systematic method for extrapolating the continuum limit of c in the two definitions is necessary.

This work was supported by the KEK supercomputing project No.02-82.

\end{document}